\documentclass{iet-njd}

\usepackage{algorithmic}
\usepackage{algorithm}
\usepackage{makecell}

\ietJournal{CMU2}
\setcopyright{open}
\ietVolume{00}
\ietYear{2021}
\ietDoi{cmu2.10001}

\received{DD MMMM YYYY}
\revised{DD MMMM YYYY} 
\accepted{DD MMMM YYYY}

\begin{document}
	
	
	\title{Towards Transactive Energy: An Analysis of Information-related Practical Issues}
	
	
	\author[af1]{Yue Chen}
	\author[af2]{Yu Yang}
	\author[af3]{Xiaoyuan Xu}
	
	\affil[af1]{Department of Mechanical and Automation Engineering, the Chinese University of Hong Kong, Hong Kong SAR, China.}
	\affil[af2]{School of Automation Science and Engineering, Xi'an Jiaotong University, Shaanxi, China.}
	\affil[af3]{Key Laboratory of Control of Power Transmission and Conversion (Shanghai Jiao Tong University), Ministry of Education, Shanghai, China.}
	
	\corresp{\textbf{Correspondence}\\ yuechen@mae.cuhk.edu.hk}
	
	\runauth{CHEN et al.}
	
	\begin{abstract}%
		The development of distributed energy resources, such as rooftop photovoltaic (PV) panels, batteries, and electric vehicles (EVs), has decentralized our power system operation, where transactive energy markets empower local energy exchanges. Transactive energy contributes to building a low-carbon energy system by better matching the distributed renewable sources and demand. Effective market mechanisms are a key part of transactive energy market design. Despite fruitful research on related topics, some practical challenges must be addressed. This review surveys three practical issues related to information exchange in transactive energy markets: asynchronous computing, truthful reporting, and privacy preservation. We summarize the state-of-the-art results and introduce relevant multidisciplinary theories. Based on these findings, we suggest several potential research directions that could provide insights for future studies.
	\end{abstract}
	
	
	\maketitle%
	
	\section{Introduction}
	\label{secI}
	
	Global climate change caused by high emissions of fossil fuels has been viewed as a critical challenge facing human society. According to the Emissions Gap Report 2021, the annual greenhouse gas emissions must be halved in eight years from 2021 to slow down the warming process effectively~\cite{unep2021emissions}. This calls for the decarbonization of energy systems through the widespread deployment of carbon-free renewable energy sources (such as wind and PV power) and electrification of energy sectors \cite{yang2021spot}. The global wind and solar installation capacity surpassed 1,500 GW by 2020 \cite{windreport2021, solarreport2021}. The number of plug-in electric vehicles (EVs) is growing at a million per year and is predicted to reach 145 million by 2030, representing 7\% of the global vehicle fleet~\cite{EVoutlook2021}. Battery energy storage (ES) is expanding, with the total installed capacity expected to reach 17GW by 2020 ~\cite{EnergyStorage2021}.
	The proliferation of distributed energy resources (DERs) is transforming power systems from a top-down, centralized to a bottom-up, decentralized structure. End users are evolving from passive consumers to proactive prosumers who can produce, use, and store energy.
	Owing to their small capacity and unstable output, DERs are currently forbidden from participating in the wholesale power market. The power grid mainly recycles excess renewable energy under a low feed-in tariff (FiT). However, this FiT scheme is unsustainable owing to marginal incentives~\cite{tushar2014three}, and high government expenses \cite{GermanyFeed-inTariffs}. As a result, the FiT scheme in Germany ended in 2021, while that in the state of Queensland, Australia, will expire in 2028 \cite{GermanyFeed-inTariffs}.  
	Although equipping renewable generators with behind-the-meter batteries can help shift the volatile renewable supply overtime to meet demand, existing technology makes it prohibitively expensive \cite{ziegler2019storage}. Transactive energy has emerged as a promising solution for empowering efficient and low-carbon energy systems by allowing DERs to exchange surplus energy locally. 
	
	Transactive energy is a relatively new concept first formally defined by the GridWise Architecture Council in 2013, which refers to ``the economic and control techniques used to manage the flow or exchange of energy within an existing electric power system with regard to economic and market-based standard values of energy" \cite{melton2013gridwise}. In particular, transactive energy provides market platforms for prosumers to exchange energy or energy services (e.g., peak demand shaving or regulation services). Prosumers can establish a coalition to exchange energy collaboratively or trade energy individually. Transactive energy is a broad concept that is known by various names in different application contexts, including energy collectives \cite{moret2018energy}, energy communities \cite{feng2019coalitional}, energy clusters \cite{chandra2021transactive}, peer-to-peer (P2P) energy trading \cite{tushar2020peer}, energy sharing \cite{chen2018analyzing}, and local electricity markets \cite{luth2018local}.
	
	\begin{figure}[t]
		\centering
		\includegraphics[width=0.9\columnwidth]{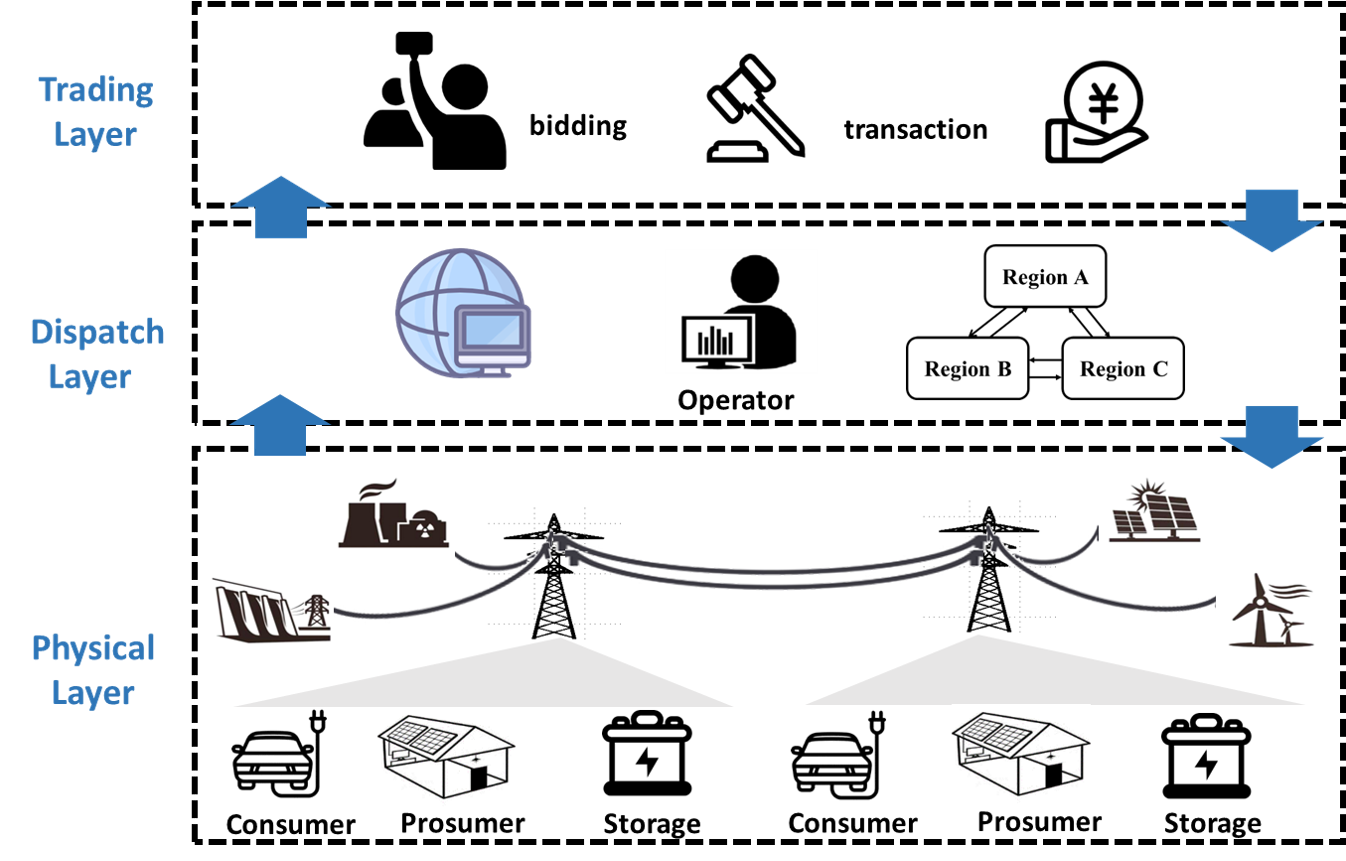}
		\caption{Conceptual model of a transactional energy system.}
		\label{fig:framework}
	\end{figure}
	
	A transactional energy system was implemented in three tiers, as shown in FIGURE \ref{fig:framework}. The physical devices and networks on which the energy transaction is performed constitute the bottom layer. As a result, the trading result must satisfy physical constraints, such as the power flow and generator capacity limits. The trading results are reviewed in the second layer to ensure this. The second layer serves as an intermediary: on the one hand, it ensures effective power delivery in the bottom layer; on the other hand, it provides information for agents' bids in the top layer. Trading-related activities appear in the top layer, where numerous individuals strategically bid their offers/demands to maximize profits. The connections between neighboring layers are supported by information and communication technologies such as communication devices, protocols, and data flows.
	
	In recent years, transactive energy has attracted significant attention. Existing work mainly focuses on three areas: \textit{market structure design}, \textit{market mechanism design}, and \textit{market implementation}. Research on \textit{market structure} studies how prosumers should be organized, and the three typical schemes are full P2P, community-based P2P, and hybrid P2P \cite{parag2016electricity}. Here, P2P is a way to achieve transactive energy, which refers to an architecture in which two individuals can interact.
	Full P2P markets allow prosumers to negotiate deals and prices directly and exclusively. In a community-based P2P market, prosumers within a community trade energy with one another to benefit the entire community. The central coordinator determines the transactions and profit allocations among prosumers inside the community. The hybrid P2P market lies between the full P2P and community-based P2P markets. More specifically, prosumers nearby form small energy communities; internally, prosumers collectively share energy to optimize the welfare of the community, and externally, the communities trade in a P2P manner. Research on the \textit{market mechanism} addresses the design of economic incentives to encourage prosumers to act towards a desired outcome. In addition to typical market design issues, the mechanism for an energy market should enable real-time energy balancing and satisfaction of network constraints. The design of a transactive energy market mechanism is dependent on market structures and communication networks, which are introduced in detail in Section \ref{secII}. The \textit{market implementation} study focuses on implementing energy transactions using advanced information and communication technologies. It is further subdivided into \textit{distributed ledger technologies}, which support secure and transparent transactions; \textit{control technologies}, which monitor prosumers' production and consumptions based on the trading results; and \textit{project demonstrations}.

	\begin{table*}[t]
		\footnotesize
		\renewcommand{\arraystretch}{1.3}
		\centering
		\caption{Summary of existing reviews}
		\label{tab:comparison-review}
		\begin{tabular}{cccccc}
			\hline 
			&  \textbf{Market structure} & \textbf{Mechanism design}  & \textbf{Ledger Technology} & \textbf{Control technique} & \textbf{Pilot project} \\
			\hline
			\cite{liu2017transactive}  & & & & \checkmark & \checkmark\\
			\cite{zia2020microgrid} & \checkmark &  & \checkmark & \checkmark & \\
			\cite{huang2021review}  & & \checkmark & \checkmark & & \checkmark\\ 
			\cite{abrishambaf2019towards}  & & \checkmark & & \checkmark & \checkmark\\
			\cite{siano2019survey}  & \checkmark &  & \checkmark & &\\
			\cite{yang2020transactive}  & & \checkmark & & &\\
			\cite{chen2017demand}  & & \checkmark & & &\\
			\cite{chen2021peer} & \checkmark & \checkmark & & & \\
			\hline
		\end{tabular}
	\end{table*}
	
	Several reviews have surveyed transactive energy from the above perspectives, as compared in TABLE \ref{tab:comparison-review}. This study focuses on the market mechanism design, an essential part of transactive energy. Although reviews such as \cite{huang2021review,abrishambaf2019towards,yang2020transactive,chen2017demand,chen2021peer} also deal with market mechanism design; they do so from a theoretical standpoint based on ideal conditions and rarely consider information-related practical concerns. In this paper, the term ``information'' refers processed, structured, and organized data. It contextualizes facts and facilitates decision-making. Many types of data (information) must be obtained or exchanged to settle transactions in a transactional energy system. Therefore, the successful transmission of truthful information while respecting privacy is essential for successful energy transactions. To the best of our knowledge, information-related practical challenges such as asynchronous computing, the honesty of prosumers, and the potential privacy leakage are crucial yet have received little attention. Owing to these practical challenges, the actual market outcome may differ from the theoretical design, resulting in a suboptimal solution. This study aims to identify critical information-related practical issues by surveying previous studies. We summarize relevant multidisciplinary theories to shed light on how these issues may be addressed. Based on these findings, we suggest future research directions. This study contributes to the smooth transition of transactive energy from theoretical design to actual execution.
	
	The remainder of this paper is organized as follows: In Section \ref{secII}, we classify the existing transactive energy market mechanisms based on the underlying market structures and communication networks, followed by three potential information-related practical issues. In Section \ref{sec-III}, we analyze each practical issue in detail by surveying related studies. Future research directions are discussed in section \ref{sec-IV2}. In Section \ref{sec-IV}, we summarize this paper.
	
	\section{Market Mechanisms}
	\label{secII}
\begin{figure*}[t]
		\centering
		\includegraphics[width=1.5\columnwidth]{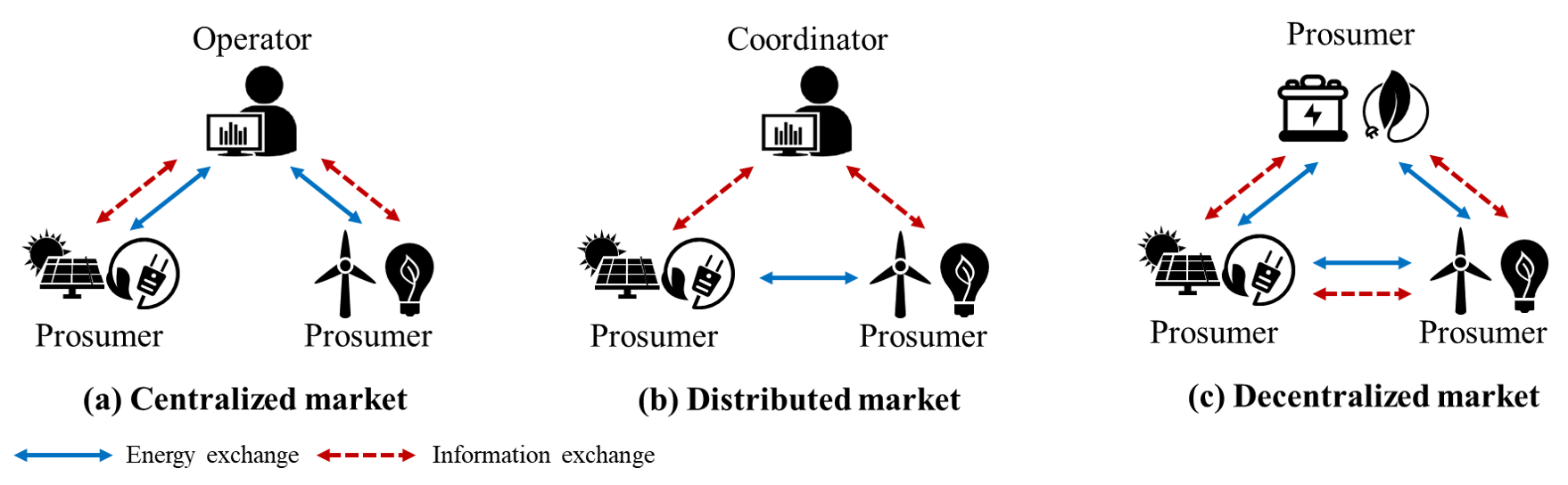}
		\caption{Three typical transactive energy markets: (a) centralized market, where both the energy transactions and information exchange happen between individual prosumer and the operator; (b) distributed market, where prosumers trade with each other and exchange information with a central entity for coordination; (c) decentralized market, where prosumers communicate and trade with each other.}
		\label{fig:market-structure}
	\end{figure*}

	Transactive energy markets aim to achieve an efficient and economical balance of energy supply and demand. They can be divided into \textit{centralized markets}, \textit{distributed markets}, and \textit{decentralized markets} according to the underlying market structures (energy exchange) and communication networks (information exchange), as illustrated in FIGURE \ref{fig:market-structure} \cite{tushar2021peer,zhou2020state}. In particular, the centralized market is a common form of community-based P2P markets, whereas decentralized and distributed marketplaces are typically built for full P2P and hybrid P2P. In a centralized market, both the transactions and information exchanges among prosumers are managed by a central operator. Prosumers are required to report their complete information (e.g., preference, renewable, and load profiles) to the central operator and to commit to the decisions made. In a distributed market, prosumers are empowered to make decisions locally. Though prosumers continue to exchange information with the central coordinator in order to achieve efficient and cost-effective energy transactions, they do not trade directly with the coordinator. In comparison, a central coordinator is completely eliminated in decentralized marketplaces, where prosumers communicate directly with one another to settle energy transactions.
	
	The coordination of prosumers' decision-making in transactive energy markets is a big difficulty due to fluctuating renewable power supplies, unpredictable demand, competing interests, varied preferences, and so on. This problem can be solved by using effective market mechanisms with well-designed price-setting and allocation criteria to guide prosumers' behaviors towards a satisfactory outcome. Since market mechanisms are built upon market structures and communication networks, in this paper, we divide the related work into the three market categories (centralized, distributed, decentralized) indicated above. We then go over their benefits and drawbacks in depth, as well as the potential information-related practical concerns.
	
	\subsection{Classification of market mechanisms}

	\begin{table*}[h]
		\renewcommand{\arraystretch}{1.3}
		\centering
		\footnotesize
		\caption{Transactive Energy Mechanisms}
		\label{tab:mechanism}
		\begin{tabular}{m{2.5cm}<{\centering}p{5cm}<{\centering}m{3cm}<{\centering}m{3cm}<{\centering}}
			\hline 
			Transactive energy markets & Models / Algorithms  & Advantages & Disadvantages  \\
			\hline
			\textbf{Centralized market} (Community-based P2P)  &        \makecell[l]{Stackelberg game \cite{liu2017energy, liu2018energy,cui2018distributed,zhang2021time,liu2018hybrid2}. \\ Coalition game \cite{liu2018hybrid,mei2019coalitional,long2019game, han2018incentivizing,han2019improving, yang2021optimal, wang2019incentive,wang2016incentivizing}.}   &  Manageable.  Maximum social benefits.  Easy to consider system constraints.   & High computation \& communication.  Suboptimal outcome under information asymmetry. Privacy issues.                                 \\
			& & & \\
			\textbf{Distributed market}  (Full P2P, Hybrid P2P) &      \makecell[l]{  Generalized Nash game \cite{wang2019integrated,chen2018energy, chen2019trading,sun2015learning,li2015demand, chen2019energy, chen2020approaching,xu2015demand, chen2021energy,chen2021flexibility}. \\ Multi-leader multi-follower game \cite{ paudel2018peer, anoh2019energy}. \\ Coordination-based dual decomposition \cite{gregoratti2014distributed, yang2018decentralized, deng2014distributed,yoo2021energy}. \\ Coordination-based ADMM \cite{baroche2019prosumer,nguyen2021distributed}.
			} & Scalable.  Low computation \& communication. Relatively easy to coordinate.  &   Communication delays. Limited flexibility of participants. Suboptimal outcome under information asymmetry. Privacy issues.                 \\
			& & & \\
			\textbf{Decentralized market} (Full P2P, Hybrid P2P) &   \makecell[l]{Generalized Nash game \cite{wang2021distributed}. \\
				Consensus-based dual decomposition \cite{rahbari2014incremental,wang2021consensus}. \\
				Consensus-based ADMM \cite{yang2019fully,yang2019distributed,rajaei2021decentralized}.
			}   & Scalable.  Low computation \& communication. High participant flexibility.  &    Communication delays. Hard to coordinate. Hard to ensure system security. Participants facing more  uncertainties. Suboptimal outcome under information asymmetry. Privacy issues.                          \\
			\hline
		\end{tabular}
	\end{table*}
	
	TABLE~\ref{tab:mechanism} compares the existing transactive energy market mechanisms. The \textit{centralized market} is typically modeled as a Stackelberg or a coalition game. In a Stackelberg game, the operator moves first to determine the energy prices with anticipation of prosumers' reactions and the prosumers follow to decide on the quantities. Prior work along this line spans applications in microgrids with renewable generators \cite{liu2017energy,liu2018energy,cui2018distributed}, electricity retail markets \cite{zhang2021time}, multi-energy systems \cite{liu2018hybrid2}, etc. In a coalition game, a centralized optimization problem is first solved to determine the buying or selling quantities of the prosumers. Then the revenue is distributed by the operator according to certain criteria. Typical criteria include the Shapley value \cite{liu2018hybrid,mei2019coalitional,long2019game} that considers each prosumer's marginal contribution, the nucleolus \cite{han2018incentivizing,han2019improving, yang2021optimal} that seeks allocations with coalition stability guarantee, the Nash bargaining method \cite{wang2019incentive,wang2016incentivizing}, etc.
	
	In a \textit{distributed market}, prosumers trade with each other rather than with the operator, although the operator will provide some signals to steer their behaviors. Generalized Nash games or multi-leader multi-follower games are commonly used to model distributed markets. In a generalized Nash game, the operator assists in settling transaction prices based on prosumers' bids but does not trade with them. The widely-applied double auction belongs to this category. In a double auction market, the potential sellers submit their ask prices, the potential buyers declare the quantities they want, and the market operator determines the clearing price. The double auction market equilibrium can be reached by solving optimizations iteratively \cite{wang2019integrated,chen2018energy} or by learning-based methods \cite{chen2019trading,sun2015learning}. Reference \cite{lian2018performance} provided a systematic procedure to evaluate the performance of double auction markets. Apart from bidding a single price or quantity, the participants can also bid functions such as the supply function \cite{li2015demand} and the generalized demand function \cite{chen2019energy}. Capacity constraints \cite{chen2020approaching,xu2015demand} and network constraints \cite{chen2021energy,chen2021flexibility} were further taken into account. In a multi-leader multi-follower game, the operator helps to match sellers to buyers \cite{ paudel2018peer, anoh2019energy}.
	In a distributed market, we may also employ optimization based mechanisms that try to solve a social optimal optimization problem in a distributed way. 
	Two well-known distributed algorithms used are the coordination-based dual-decomposition and the coordination-based alternating direction method of multiplier (ADMM). Based on the dual-decomposition algorithm, the optimal energy trading schedules between microgrids \cite{gregoratti2014distributed}, EVs \cite{yang2018decentralized}, and between utility company and users \cite{deng2014distributed,yoo2021energy} were given. The dual decomposition method is straightforward, however, its convergence relies on a set of strong assumptions \cite{boyd2011distributed}, which may not hold for engineering problems. By adding a regularization term, the ADMM method benefits from fast convergence, broad scalability, and good stability (can adapt to different communication sparsity). Several peer-to-peer energy trading mechanisms were developed based on the ADMM method with a coordinator \cite{baroche2019prosumer,nguyen2021distributed}.
	
	\textit{Decentralized markets} differ from distributed markets in that they run without a central coordinator. For example, an energy sharing mechanism based on information exchange between neighbours was proposed in \cite{wang2021distributed}. A consensus-based variant of the dual decomposition method and the ADMM method were applied in \cite{rahbari2014incremental,wang2021consensus} and \cite{yang2019fully,yang2019distributed,rajaei2021decentralized}, respectively.
	
	\subsection{Advantages, disadvantages, and challenges}
	\label{secII-2}
	The pros and cons of the three market categories (centralized, distributed, decentralized) are summarized in TABLE \ref{tab:mechanism}.
	
	\textit{Centralized market} is more manageable, can easily take into account network constraints, and can maximize the social welfare. However, centralized markets rely on a powerful central entity to make decisions for all the prosumers involved. The prosumers are obligated to provide the system operator with complete information (e.g., load and renewable output profiles). As a result, the implementation of a centralized market is frequently hampered by high computation and communication burdens. \textit{Distributed markets} and \textit{decentralized markets}, on the other hand, benefit from low/reduced communication and computing complexity because the main processing tasks are distributed to prosumers. Please note that here “high” or “low” communication refers to the communication overhead in each iteration constrained by the channel capacity. In a centralized market, all participants need to submit complete information to the operator, including but not limited to their utility functions, demand data, etc., while in a distributed or a decentralized market, the participants usually only need to submit a bid each time. Furthermore, in a \textit{distributed market}, it'd be relatively easy to coordinate the participants, but meanwhile, the flexibility of participants is limited to some extent since they are still influenced by the operator. In a \textit{decentralized market}, the participants have higher flexibility, but their behaviors can be harder to coordinate due to the lack of a central operator. This could result in a social welfare far from optimum. Moreover, the P2P energy trading outcome is unpredictable for the system operator, so ensuring system security is difficult. Moreover, without a coordination, the participants are facing more uncertainties.
	
	In addition, one potential challenge of the distributed/decentralized market is the communication hazards among prosumers. To be specific, normally, an iterative process is adopted for coordinating prosumers in a distributed or decentralized market. An unstable communication network can cause communication delays or losses. However, most existing transactive energy mechanisms in distributed and decentralized markets are based on synchronous computations, which means that during the resolution process, each prosumer must wait for information from all its interconnected partners. When facing long communication delays or communication losses, this synchronization manner may be impracticable due to the calculation timeouts.
	
	Furthermore, existing mechanisms usually assume a symmetric information structure, which means that different participants have the same set of information about the same thing/parameter. This assumption is far from practice, for example, network constraints are only available to the operator, whereas capacity constraints are known to each participant. Some participants may deliberately misrepresent their private information to mislead others, which is referred to as ``asymmetric information''.
	
	Lastly, all three types of markets (centralized, distributed, and decentralized) rely on information flow among market participants (operators and prosumers), hence privacy protection is crucial. We may need to expose some individual statistics to the public in order to facilitate the energy transaction. Even if we do not divulge the sensitive information directly, an opponent might be able to reverse-engineer the sensitive data from what we have released. For example, a prosumer's living habits (staying or leaving home) may be learned from their released load profiles \cite{chen2015significant, schweizer2015using}. Such potential privacy breaches may deter prosumers from entering the transactive energy market.
	
	Despite the fruitful work on transactive energy market mechanism designs, the foregoing practical concerns (communication delays, asymmetric information, and privacy leakage) associated with information exchanges have not been adequately addressed. Market mechanisms that permit asynchronous computation, incentivise participants to report truthfully, and protect participants' privacy are required to fix these practical issues and to facilitate the introduction of transactive energy markets. In the following, we survey the existing work addressing such practical difficulties and introduce relevant multidisciplinary theories. Then, we suggest several potential research directions for future investigations.

	\section{Information-related Practical Issues}
	\label{sec-III}
	
	In this section, we will discuss the three practical issues (i.e., asynchronous computing, truthful reporting, and privacy preserving) in detail. For each practical issue, we begin with an introduction to its definition and challenges; next we summarize relevant theories; finally, we outline the current efforts in energy markets and the remaining difficulties. 
	The key points are highlighted in TABLE \ref{tab:practical-issues}.
	\begin{table*}[t]
		\setlength\tabcolsep{1.5pt}
		\renewcommand{\arraystretch}{1.3}
		\centering
		\footnotesize
		\caption{Summary of Information-related Practical Challenges}
		\label{tab:practical-issues}
		\begin{tabular}{m{2.0cm}<{\centering}m{3cm}<{\centering}m{3.0cm}<{\centering}m{2.5cm}<{\centering}m{2.5cm}<{\centering}m{2.5cm}<{\centering}}
			\hline 
			Practical Issues & Description & Theories	& Limitations & Existing work & Future Directions\\
			\hline
			\makecell{Asynchronous \\computing} & Communication delays, losses, and solver diversities lead to long or invalid negotiations.  & Asynchronous distributed algorithms, e.g., flexible ADMM  \cite{hong2016convergence, wei2013}, asynchronous ADMM \cite{zhang2014asynchronous, chang2016}.  & Bounded communication delays. 
			One-way communication delays. & \cite{moret2018negotiation, dong2021convergence,huang2008asynchronous} & Extensions to consider the realistic communication delay models. Trade-off between waiting time and convergence speed.  \\
			& & & & & \\
			\makecell{Truthful \\ reporting} & One entity has more or better information than the other, causing imbalance of power in transactions. & Principal-agent framework \cite{AsInf-R2}.  Myerson's Lemma \cite{myerson1981optimal}.  &  Rely on simple models.  Influence of prosumer's preference and attitude towards risks not fully studied. & \cite{qiu2017risk, abbink2003asymmetric, weidlich2008critical, chen2018optimal,leong2019auction,chen2021pricing} & Try to incorporate network constraints. More accurate prosumer models by data-driven methods.\\
			& & & & & \\
			\makecell{Privacy \\ preserving}& An adversary might be able to reverse-engineer the sensitive data from what we have released. &  Homomorphic-encryption \cite{armknecht2015guide}. Differential-privacy \cite{dwork2014algorithmic}  & Only support certain kinds of operations, or face high computational burden.  Complexity of theoretical analysis. & \cite{yi2021energy, son2020privacy,gaybullaev2021efficient, samuel2021secure,thandi2021privacy, li2019towards, li2011practical} & Analyze which data may reveal privacy. Tradeoff between privacy and security.\\
			\hline
		\end{tabular}
	\end{table*}

	\subsection{Asynchronous computing}
	As discussed in Section \ref{secII-2}, full P2P markets (including distributed and decentralized markets) are effective alternatives to the centralized counterpart due to lower computation and communication costs, greater prosumer flexibility, and the ability to preserve privacy to some level. To achieve a desired overall outcome with local decision-making, prosumers need to be coordinated through an iterative process supported by recurring communications. Conventionally, synchronous algorithms are adopted, in which the computing agents will move to the next iteration only after the information of all its interconnected agents are received.
	However, due to the geographic dispersion of participants and the differences in local computing devices, synchronous algorithm based full P2P markets may face communication hazards. For example, communication delays induced by unreliable communication networks may result in a large increase in the processing time of synchronous algorithms, rendering them ineffective. Worse, the entire P2P market may get stalled as a result of information loss in communication.
	
	Relevant studies that can provide insights on this issue are the asynchronous distributed algorithms for constrained optimization as summarized in TABLE \ref{tab:asynchronous_distributed_approaches}. Currently, there is no unifying definition of "asynchronous" and the two common forms of " asynchrony" are: 1) the situation in which each computing agent is activated with some probability or rules at each iteration. A typical algorithm is the flexible ADMM developed in \cite{hong2016convergence, wei2013} that can achieve the centralized optimal solution in a distributed manner while allowing each agent to be activated with some probability or at least once during a given length of period. This is different from the classic synchronous ADMM, in which all agents are activated at each iteration. The flexible ADMMs often assume updated and synchronized information among the network, which means that the possible communication delays are neglected.
	2) the situation in which there are communication delays. A typical algorithm is the asynchronous ADMM proposed in \cite{zhang2014asynchronous, chang2016} that takes into account the communication delays and allows an asynchronous pace of local computations. To be specific, each agent starts the local calculation as long as a certain number of messages from its interconnected agents is received. This is different from the traditional synchronous ADMM in which each agent will not start the computation until all messages from its interacted agents arrive. The asynchronous ADMMs often assume a bounded communication delay, which indicates that the information from interconnected agents can be at most $\tau$ old.
	
	When it comes to transactive energy markets, reference \cite{moret2018negotiation} showed that both the communication and computation delays will lead to oscillations of the negotiation process when using the traditional synchronous ADMM, resulting in a longer negotiation process. To overcome this issue, a novel design of peer-to-peer energy trading market based on the asynchronous ADMM algorithm in \cite{zhang2014asynchronous} was developed \cite{dong2021convergence}. To be specific, when a certain number of messages are received, a participant can progress to the next iteration without waiting for the information from all its trading partners.  The proposed market in \cite{dong2021convergence} was shown to be more efficient due to the reduced waiting time. Asynchronous algorithm was also applied in \cite{huang2008asynchronous} to facilitate the energy trading among interconnected
	electricity markets.
	
	In general, the studies in this line still face some challenges. \textit{First}, as discussed above, communication delay is ignored in the flexible ADMMs, and although it is considered in the asynchronous ADMMs, the delay must be within a certain range. As a result, before using such methods to create asynchronous transactive energy markets, we must ensure that the communication latency assumptions are met. \textit{Second}, existing algorithms only address one-way communication delays (often the delays from local computing agents to the central coordinator). However, in transactive energy markets, two-way communication delays can happen.
	
	
	\begin{table*}[t]
		\setlength\tabcolsep{2pt}
		\renewcommand{\arraystretch}{1.3}
		\centering
		\footnotesize
		\caption{Asynchronous distributed optimization algorithms}
		\label{tab:asynchronous_distributed_approaches}
		\begin{tabular}{m{2.5cm}<{\centering}m{3cm}<{\centering}m{3cm}<{\centering}m{2.5cm}<{\centering}m{2.5cm}<{\centering}m{2.5cm}<{\centering}}
			\hline 
			Algorithms & Asynchronous Features & Main assumptions   & Benefits & Limitations	& References \\
			\hline
			Flexible ADMM        &   Each computing  agent is activated with some probability or by some rules in each iteration.    &      Each agent is selected with a positive probability, or  at least once during a given length of period.          & Reduce  the communication and communication overhead.  &   Ignore communication delays. & \cite{hong2016convergence, wei2013} \\
			& & & & &\\
			Asynchronous ADMM & Each  agent starts local computation with   a certain number of messages.      &       Bounded communication delays.        &  Reduce the computation and communication overhead. &  Depend on the bound of delays; only consider
			one-way communication delays. & \cite{zhang2014asynchronous, chang2016} \\
			\hline 
		\end{tabular}
	\end{table*}
	
	\subsection{Truthful reporting}
	In practice, the adverse impact of neglecting information asymmetry is another critical issue we cannot ignore. It has been observed that in the Day-Ahead Load Response Program (DALRP) in New England, U.S., most of the customers tended to offer lower day-ahead load reduction levels than what they are actually capable of, in order to earn money by making use of the difference between day-ahead and real-time locational marginal prices (LMPs) \cite{AsInf-R2}. Basically, there are two types of asymmetric information: adverse selection (ex-ante asymmetry) and moral hazard (ex-post asymmetry) \cite{laffont2013theory}. Adverse selection refers to a situation in which the sellers have certain information of a product but the buyers do not have. This kind of asymmetric information happens \textit{before} transactions, and so is called "ex-ante asymmetry". The most well-known example of adverse selection is the "lemon market" \cite{akerlof1978market}. The aforementioned case in New England is another example. Moral hazard refers to the idea that an agent protected from risk will behave differently than if they were not protected. This kind of asymmetric information happens \textit{after} transactions, and so is called "ex-post asymmetry". 
	Despite the fact that asymmetric information problems have been extensively studied in economics, few research in energy markets have addressed these concerns due to the complexity imposed by the network constraints. Finding tools to analyze the consequences of asymmetric information and implementing truth-telling mechanisms are critical to overcoming this challenge.
	
	Reference \cite{qiu2017risk} discovered that households that consume less energy during peak hours are more likely to join the demand response program, caused by adverse selection. Reference \cite{abbink2003asymmetric} found that uniform pricing is more efficient than discriminatory pricing due to the presence of asymmetric information, which contradicts the common belief that discriminatory pricing is better. The significance of taking information asymmetry into account in the retailer's optimal pricing problem has been highlighted \cite{weidlich2008critical}. The above studies reveal the necessity of considering information asymmetry. One important technique is the "principal-agent framework" in contract theory \cite{laffont2013theory}. The asymmetric information structure between the retailer and consumers is modeled as a signal game, and the optimal contract of energy mix was proposed to tackle this problem \cite{chen2018optimal}. Another important technique is the Myerson's Lemma in auction design \cite{myerson1981optimal}. A famous implementation of the Myerson's Lemma is the Vickrey-Clarke-Groves (VCG) mechanism, which was used to limit power losses in P2P energy trading \cite{leong2019auction}. Besides, there are some heuristic designs to encourage truth-telling. For example, a temporal locational marginal pricing scheme was proposed to  offer price-taking firms with incentives to bid honestly \cite{chen2021pricing}. 
	
	In general, developing truth-telling mechanisms is difficult, and there is no cohesive theory for such designs at this time. The ``principal-agent'' framework and the Myerson's Lemma may give some insights, but their models are relatively simple and still far from reality.
	
	
	\subsection{Privacy preserving}
	Although distributed and decentralized markets can preserve privacy to some extent by limiting the amount and frequency of exchanged data, they may still suffer from potential private information leakage. Typical approaches for protecting data privacy are the homomorphic-encryption based methods \cite{armknecht2015guide} and the differential-privacy based methods \cite{dwork2014algorithmic}.
	The \textit{homomorphic-encryption} (HE) based methods first encrypt the sensitive data, outsource and do the computation while encrypted, and finally decrypt the result. It can assure that the decrypted computation results are the same as those obtained using the unencrypted data. The HE-based method aims to be noise-free so that we can get the same market clearing outcome as the original market. However, the noise-free methods can only support certain kinds of operations on the ciphertext. For example, the RSA cryptosystem \cite{al2012enhanced} can be used for an unbounded number of modular \textit{multiplications}, while the Paillier cryptosystem \cite{paillier1999public} can be used for an unbounded number of modular \textit{additions}. Fully homomorphic encryption schemes such as the Bootstrapping \cite{abney2002bootstrapping} were developed, but they might be computationally intensive.
	The \textit{differential-privacy} (DP) based methods preserve privacy by adding some controlled noises (e.g. Laplace noises) to the inputs or outputs. The concept of $\epsilon$-privacy was proposed \cite{dwork2006differential} to quantify the privacy loss associated with any two datasets differing on a single element. The DP-based method is easy to implement, but the increased noise reduces accuracy. Moreover, noise perturbation may lead to divergence of the iterative process in a distributed or decentralized market.

	The HE-based methods are often combined with blockchain technologies \cite{yi2021energy}. The inner product functional encryption method was used in \cite{son2020privacy,gaybullaev2021efficient} for the energy trading in a blockchain-embedded smart grid. The additive homomorphic encryption were applied in \cite{samuel2021secure,thandi2021privacy}.
	There are relatively few work applying the DP-based methods in energy trading due to its theoretical complexity. A novel privacy-preserving online double auction
	scheme was proposed in \cite{li2019towards} to facilitate the energy trading among electric vehicles.
	Besides the HE-based and the DP-based methods, an anonymous auction protocol was developed in \cite{li2011practical}.
	
	In general, the existing HE-based or DP-based methods have limited application scenarios. Despite some newly developed fully homomorphic encryption schemes, the computational burden and accuracy in the long run are still challenging. Moreover, distributed ledger technology and blockchain are important tools for preserving privacy in transactive energy markets, but they are more related to the implementation part, so we do not go into detail in this paper. Readers who are interested can refer to \cite{wang2019energy,siano2019survey}.
	
	
	\section{Future Research}
	\label{sec-IV2}
	According to the above discussions, we propose the following future research directions for each information-related issue.
	
	(1) \textit{Asynchronous-computing}. The present research solely analyzes unidirectional communication delays that must be within a certain range. As a result, effective algorithms that account for more realistic situations, such as two-way delays, are vital, albeit difficult. Furthermore, it is worth mentioning that asynchronous distributed algorithms often accelerate an iteration using partial information at the expense of slowing convergence. The tradeoff between waiting time in each iteration and time required to converge is an important problem that has received little attention so far. Therefore, it's worth investigating how market mechanisms with asynchronous behaviors should be designed to maximize overall computational efficiency.
	
	(2) \textit{Truthful-reporting}. Because energy transactions take place on energy networks, they must adhere to network restrictions. Despite the different economic theories for dealing with the problem of asymmetric information, such as the "principal-agent" framework and Myerson's Lemma, they are all based on simple individual constraints. Hence, how to incorporate the complicated coupling physical constraints is still an open question. Furthermore, the agent's risk preference and attitude, i.e., whether it is risk-averse, risk-neutral, or risk-appetite, may influence the mechanism design. This can be solved by data-driven approaches.
	
	(3) \textit{Privacy-preserving}. There are two challenges to fully ensuring privacy in a transactive energy market: 1) the design of privacy-preserving algorithms. 2) Implementation in energy markets. For the former issue, the computational burden of the HE-based method and the theoretical intricacy of the DP-based method are still obstacles to overcome. For the latter issue, in most energy trading scenarios, we do not directly communicate sensitive data (e.g., load data), but rather some bids. Whether these bids may divulge private information needs investigation. How to protect privacy by encryption or adding some noises without violation of security constraints remains to be studied.

	\section{Conclusion}
	\label{sec-IV}
	This paper covers three information-related practical issues in designing transactive energy market mechanisms: asynchronous computing, truthful reporting, and privacy preservation. We outline the most recent findings and related multidisciplinary theories. The key takeaways are summarized below:
	
	\begin{itemize}
		\item \textit{Asynchronous-Computing}. Distributed algorithms for constrained optimizations such as the flexible ADMM and the asynchronous ADMM can help solve information-related practical issues. Interesting future research directions include removing the stringent assumptions in existing algorithms and improving the tradeoff between the waiting time in each iteration and convergence time.
		\item \textit{Truthful-Reporting}. Mechanism design methods based on the "principal-agent" framework and Myerson's Lemma provide insights into this issue. However, incorporating complex physical constraints in a transactive energy market is a major concern.
		\item \textit{Privacy-Preserving}. There are two main approaches to tackle this challenge based on homomorphic encryption and differential privacy. The high computational burden, theoretical intricacy, and limited application in energy markets must be investigated.
	\end{itemize}
	
	\section*{Acknowledgements}
	This work was supported by the CUHK Direct Grant for Research No. 4055169.
	
	\printorcid
	
	\balance
	
	\nocite{*}
	\bibliography{iet-njd}
	\bibliographystyle{iet}
	
\end{document}